\documentclass{aa}
\usepackage{epsfig} 
\usepackage{amsmath}
\usepackage{amssymb}
\usepackage{array}
\usepackage{natbib}

\newcommand{\myref}[1]{~\hspace{0pt plus 1pt minus 1pt}\ref{#1}}

\newcommand{\sectref}[1]{Sect.\myref{#1}}

\newcounter{saveqn}
\newcommand{\alpheqn}{\refstepcounter{equation}\setcounter{saveqn}{\value{equati
on}}%
\setcounter{equation}{0}%
\renewcommand{\theequation}{\mbox{\arabic{chapter}.\arabic{saveqn}\alph{equation
}}}}
 
\newcommand{\reseteqn}{\setcounter{equation}{\value{saveqn}}%
\renewcommand{\theequation}{\arabic{chapter}.\arabic{equation}}}

\newcommand{\curl}{\operatorname{curl}}
\newcommand{\dv}{\operatorname{div}}

\renewcommand{\exp}{\operatorname{exp}}

\newcommand{\itover}[2]{\,\hspace{.3mm}#1{\!\hspace{-.3mm}#2}}%verallgemeinern!!

\newcommand{\itdot}[1]{\itover{\dot}{#1}}

\newcommand{\Bvec}{\vec{B}}
\newcommand{\Evec}{\vec{E}}
\newcommand{\bvec}{\vec{b}}
\newcommand{\jvec}{\vec{j}}
\newcommand{\keV}{{\mbox{k$\hspace{.3mm}$}\mbox{e$\hspace{-.1mm}$V}}}

\begin{document} 
 
\title{Spot--like Structures of Neutron Star Surface Magnetic Fields} 

\author{U. Geppert \inst{1}, M. Rheinhardt \inst{1} \and J. Gil \inst{1,2}}  
\offprints{U. Geppert, \email{urme@aip.de}}
 
\institute{Astrophysikalisches 
Institut Potsdam, An der Sternwarte 16, 14482, Potsdam, Germany \and 
Institute of Astronomy, University of Zielona G\'ora, Lubuska 2, 65--265 
Zielona G\'ora, Poland} 
 
\date{Received date ; accepted date} 
 
\abstract{
There is growing evidence, based on both X--ray and radio observations of
isolated neutron stars, that besides the large--scale (dipolar) magnetic field, 
which determines the pulsar spin--down behaviour, small--scale poloidal field 
components are present, which have surface strengths one to two orders of 
magnitude larger than the dipolar component. 
We argue in this paper that the Hall--effect can be an efficient process in 
producing such small--scale field structures just above the neutron star 
surface.
It is shown that due to a Hall--drift induced instability, poloidal 
magnetic field structures can be generated from strong 
subsurface toroidal fields, which are the result of either a dynamo or 
a thermoelectric instability acting at early times of a 
neutron star's life.
The geometrical structure of these small--scale surface
anomalies of the magnetic field resembles that of some types of ``star--spots''.
The magnetic field strength and the length--scales are comparable with values
that can be derived from various observations.
}
\authorrunning{Geppert, Rheinhardt, Gil}
\titlerunning{Spots in Neutron Star Surface Magnetic Fields} 
\maketitle 
 
\section{Observational evidence}
\label{obs} 
There exists both observational and theoretical evidence of the existence of 
strong small--scale magnetic field structures in the surface layers of isolated 
neutron stars.
More and more, it becomes clear that neutron stars of nearly all 
ages must posses magnetic field structures, which are much more complicated than 
the simple assumption of a star centered magnetic dipole suggests.
The polar surface magnetic 
field strength of such a dipole is conventionally estimated by 
$B_d \sim 6.4 \cdot 10^{19} \sqrt{P\dot{P}}$~G, where standard neutron star 
quantities are assumed and the rotational period $P$ (in seconds) and its 
temporal derivative $\dot{P}$ are taken from radio and/or X--ray timing
observations.
 
Recently, \citet{BSP03} reported the (marginal) detection of an 
electron cyclotron line at about $3.3$~\keV~in the Chandra--spectra of the 
millisecond pulsar \object{B1821--24}.
The energetic location of that line corresponds to a surface 
magnetic field strength of $3\cdot 10^{11}$~G, while $P$ and $\dot P$ yield a 
dipolar surface field strength of about $4.5\cdot 10^9$~G, almost two orders of
magnitude lower.

\citet{HSH03} analysed the XMM--spectra of the isolated 
neutron star \object{RBS 1223} and found evidence for a proton cyclotron absorption line 
in the energy range of $0.1\ldots0.3$~\keV, which is consistent with a magnetic 
surface field strength of $(2\ldots6)\cdot 10^{13}$~G.
While $P=10.31$~s seems to be a 
well settled value for the rotational period, the determination of $\dot{P}$ is 
a much more delicate task in this case.
Recent evaluations of XMM data indicate, however, that the 
global dipolar field attributed to $P$ and $\dot{P}$ is at least one order of 
magnitude smaller than the field corresponding to the proton cyclotron line 
(Schwope 2003, private communication).
Likewise, for the pulsar \object{1E 1207.4-5209} a surface field of $\sim 1.5\cdot 10^{14}$ G
has been estimated \citep{S02}, whereas the estimate for its dipolar 
field is $(2\ldots4)\cdot10^{12}$ G \citep{P02}.

These X--ray observations indicate that, apart from a large--scale 
dipolar magnetic field which determines the spin--down behaviour 
of the pulsar, much stronger but short--ranged field components 
close to the neutron star surface exist.
They do not 
affect significantly the braking of the star's rotation, but can affect the
magnetospheric processes in the vicinity of the polar cap
(defined by the totality of all open dipolar field lines). 

%An independent evidence of strong nondipolar surface magnetic fields 
%comes from pulsar radio emission. First (WHAT IS SECOND?), 
It is commonly accepted that  
pulsar radio emission is generated within a dense  
electron--positron plasma, the creation of which 
requires an  
ultra--strong potential drop that accelerates charged particles  
along curved magnetic field lines. The observed phenomenon of drifting 
subpulses 
strongly suggests that this potential drop results from the deficiency 
in the actual charge density with respect to the so--called co--rotational 
charge density \citep{GJ69} just above the polar cap surface 
\citep[see][]{GMG03}. %for a review 
The formation of such a charge--depleted region (called `polar gap' after 
\citet{RS75}, who proposed it for the first time) 
requires a strong and highly non--dipolar surface magnetic field,  
with radii of curvature much smaller than $10^6$ cm 
and magnitudes close to $10^{13}$ G, irrespective of the dipolar field strength 
inferable from the pulsar spin--down \citep{GM01,GM02,GMG03}. 
Moreover, several periodicities observed in the phenomenon of drifting 
subpulses strongly suggest that the subpulse--associated plasma filaments  
\citep[called `sparks' after][]{RS75} circulate around a  
local magnetic pole\footnote{
A surface region with a magnetic flux prevailingly normal to the surface and
with a certain
degree of axisymmetry.} \citep{DR99,DR01,GS03}.
%I THINK THAT THE TERM "NORMAL FLUX", WHICH I BELIEVE MEANS PERPENDICULAR TO THE SURFACE,
%IS MISSLEADING. I BELIEVE THAT THIS EXPLANATION IS NOT NECESSARY AND WE SHOULD DROP THE TEXT
%BETWEEN XXXXXXOR MODIFY IT TO AVOID CONFUSION: NORMAL - DIPOLAR?
Accordingly, the small--scale surface magnetic field  
anomalies are supposed to show spot--like structures allowing a persistent arrangement of drifting 
sparks (due to the well known $\Evec \times \Bvec$ plasma drift mechanism)  
in the form of quasi--annular patterns \citep{GS00,FCM01}.

\section{The Hall--drift induced instability} 
 
It is well known that the Hall--effect, probably via a cascade 
\citep{GR92}, 
causes the generation of smaller scaled magnetic field 
components out of an existing large scale field. 
This process has been discussed for neutron star magnetic fields by a 
number of authors \citep[see, e.g.,][]{SU97,VCO00,HR02}.
Here, however, we want to refer to the fact that small--scale poloidal fields 
close to the neutron star surface can be generated from a 
subsurface toroidal magnetic field by a Hall--drift induced instability (HDI).
The basic prerequisite for this instability is that a 
sufficiently strong {\em and} inhomogeneous background field exists 
(\citet{RG02,GR02} and \citet{RKG03}[RKG03]).
We demonstrate, that strength and spatial structure of the HDI
modes resulting from a realistic NS crust model are consistent with quantities
derived from observations.

The decay of a magnetic field in the almost crystallised crust of a NS is
governed by
\begin{equation}
\begin{aligned}
\phantom{.}&\itdot{\Bvec} = -\curl\big(\,\eta\curl\Bvec+\alpha(\curl\Bvec\times\Bvec)
\,\big)\\
\phantom{.}&\dv \Bvec = 0\; ,
\end{aligned}\label{indeq}
\end{equation}
where the diffusivity $\eta$ and the Hall--parameter $\alpha$ determine the
intensity of Ohmic decay and Hall--drift, respectively, in their spatial
distribution. % (for details see RKG03).
Linearization of
Eq.~(\ref{indeq}) with respect to a background field $\vec B_0$ yields
\begin{equation}
\begin{aligned}
\phantom{.}&\begin{aligned}
  \dot{\bvec} = &-\curl\,(\eta\,\curl\bvec) \\
              &-\curl\big(\alpha\,(\,\curl\Bvec_0
                 \times\bvec+\curl\bvec\times\Bvec_0\,)\big)
 \end{aligned}\\
\phantom{.}&\dv \bvec = 0 
\end{aligned}\label{indeqlin}
\end{equation}
describing the evolution of a small magnetic perturbation $\vec b$.
For conditions as realised in the crusts of cooling isolated neutron stars, 
unstably growing perturbations
$\vec b = \hat{\vec b}(\vec r) \exp{\gamma t}\,,\;\gamma > 0\,$, can exist, with
typical growth times $\gamma^{-1} \sim 10^{4...5}$ years.

The existence of this instability has been shown in a slab geometry,
approximating the neutron star crust geometry locally.
Typical crustal density profiles were adopted.
Boundary conditions are defined by assuming a perfect conductor being adjacent
to the slab at the bottom and vacuum being adjacent at the top.
That is, corresponding to the superconductivity of the NS core,
the magnetic and electric fields do not penetrate under the
bottom of the slab 
whereas the magnetic field is continued above the slab
as a potential field approximating the conditions outside the NS.

The background fields $\Bvec_0$ we used are parallel to the slab; they vary strongly  with
depth and vanish at the bottom of the slab.
Fields of that kind, especially toroidal ones,
may exist as relics of a short, but very efficient convective dynamo phase 
acting in the proto--neutron star \citep{TD93,UG03}. 
However, it is perhaps more likely that such background fields are due to the 
action of a thermoelectric instability, which amplifies toroidal 
seed fields very effectively and sets on compulsory when $T_{\mathrm{s6} 
}^4/g_{\mathrm{s14}} \ga 100$ (\citet{WG96}, $T_{\mathrm{s6}}$ -- surface
temperature in $10^6$~K, $g_{\mathrm{s14}}$ -- surface gravity in
$10^{14}$~cm~s$^{-2}$). 
Such conditions are met in almost all newly born neutron stars and 
are maintained up to an age of $\sim\!1000$ years. 
Moreover, when a strong temperature gradient is 
established again by  any external or internal process, the thermoelectric 
instability will be switched on again. It can thus, at later stages too, 
produce a toroidal background field 
which is capable of providing the conditions for the onset of 
the HDI.
Hence, the initial strength of the background  field is 
determined essentially either by the vigour of the dynamo or by the strength of 
the temperature gradient and may well exceed $10^{14}$~G locally inside the
crust. 

Of course, in any real situation the background field is not exactly
parallel to the slab: Within the region we are interested in,
the polar cap, a normal background field arises quite naturally. Its influence
on the properties of the HDI can at the moment only be extrapolated from
the special case of a {\em homogeneous} normal background field component in a homogeneous slab. 
Then, the tangential part of the background field
rotates about a normal axis with an angular velocity proportional to
the normal field component. The HDI continues to exist if the ratio of normal to tangential
background field components is not too large.
%Its dispersion relation depends no longer on the individual components
%of the wave number vector, but on its modulus.
If the normal field is small,
what we want to assume here, the background configuration rotates slowly and we expect no major
differences in the perturbation modes in comparison with those obtained for
completely tangential background fields to appear.

The HDI is described in detail for 
conditions realised in the crusts of isolated cooling neutron stars in
RKG03 and we refer the reader to 
that extended presentation.
Its main result is that the HDI occurs for a variety
of realistic crust conditions and that the obtained growth times
are short enough to cause observable consequences. 
%competition between the 
%temperature and density dependent ohmic decay and the Hall--effect, which is 
%as well density and in addition background field dependent, determines the
%spatial structure and the growth rate of the perturbation. 

The easiest way to get an idea on how the HDI acts is considering its analogy
to field generation by hydrodynamic dynamo action. One major ingredient of dynamo models,
e.g., those
explaining the solar magnetic field, is shear motion (in axisymmetry typically occurring as
differential rotation) which is capable of
generating strong toroidal (in axisymmetry: azimuthal) fields from weak poloidal
(in axisymmetry: meridional) fields by `winding up'. In our context, the motion of the electron
fluid, in which the field is partially frozen in, is well able to provide shear (see the term 
$\alpha\,\curl\Bvec_0 \times\bvec$ in \eqref{indeqlin}).
To get a successful dynamo, shear has to be completed by another effect
generating in turn poloidal fields from toroidal ones. Mean--field dynamo theory
has identified various possibilities for that, amongst which one (the so--called $\vec{\omega}\times\jvec$ effect)
has exactly the same mathematical structure like the second Hall term $\alpha\curl\bvec\times\Bvec_0$
in \eqref{indeqlin} \citep[see, e.g.,][]{R69}. Thus, a cycle of mutual amplification of poloidal and toroidal fields can 
establish resulting in a growing total field. 
%For a qualitative illustration of this cycle we refer the reader to the Appendix of RKG03. 
 
\section{The small--scale poloidal field} 
 
\begin{figure} 
\begin{center}
\epsfig{file=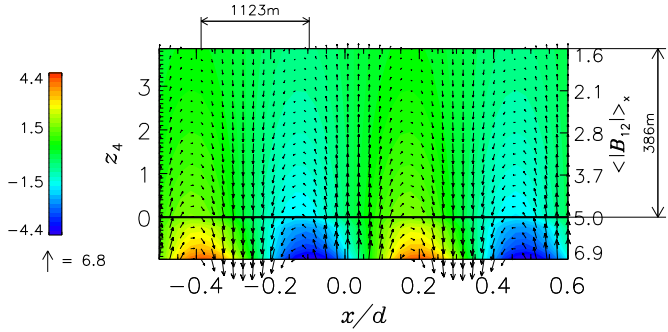, width=.98\linewidth}\\*[1mm] 
\caption{\label{eigfunc1} Structure of a typical magnetic field
perturbation generated by the Hall--drift induced instability from a 
toroidal crustal background field having a maximum field strength of 
about $10^{14}$~G.
The growth time of the perturbation is of the order of 
$5\cdot10^4$ years.
The neutron star model behind relies on the Pandharipande--Smith equation
of state, its crustal thickness $d\sim3800$~m.
The vertical scale $z$ is arbitrarily limited to
$z=d/10=386$~m, where the magnetic field strength is
reduced by a factor of about $3$ compared with its surface value.
The distance between adjacent spots in 
$x$--direction $\sim 1.1$~km, in $y$--direction $\sim 3$~km
(see Fig.~\ref{eigfunc2} for that dimension).
Colour encoding corresponds to the normal
field component, $<\cdot >_x$ denotes the average with respect to $x$.
All magnetic field values are given in units
of $10^{12}$~G, $z_4$ is in units of $10^{4}$~cm} 
\end{center}
\end{figure}

\begin{figure}
\begin{center}
\hspace*{-1.2cm}%
\epsfig{file=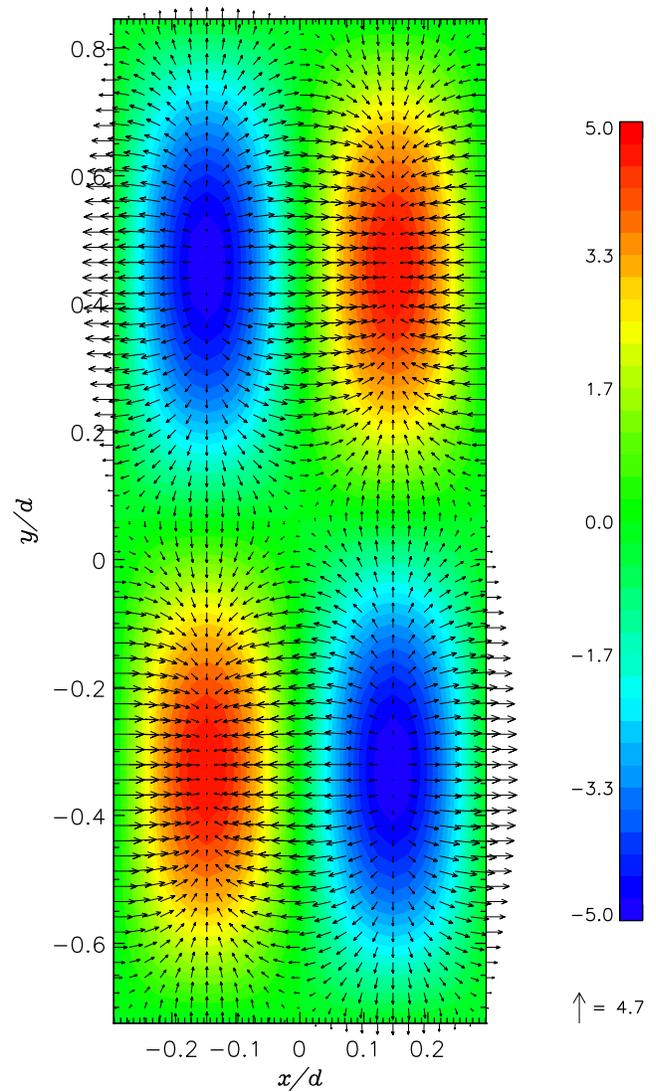, width=1.15\linewidth}\\*[-1mm] 
\caption{\label{eigfunc2} The same magnetic field perturbation as in
Fig.~\ref{eigfunc1} but viewed from
above and zoomed into a smaller $x$ interval.
Note, that this field has to be superimposed on the global dipolar
field having its pole in the vicinity of one of the spot centers
to get the actual open magnetic field lines.} 
%, necessary for particle acceleration.} 

\vspace{-6mm}
\end{center}

\end{figure}

The structure of a typical\footnote{Pandharipande--Smith equation
of state, initial penetration
density $\rho_0=10^{13}$~g~cm$^{-3}$, cubic background field $\Bvec_0$
with initial polar field strength $10^{14}$~G,
model age $3 \cdot 10^5$ years.} unstable (i.e., exponentially growing) perturbation
field (or eigenmode) is shown in Figs.~\ref{eigfunc1} and \ref{eigfunc2} 
(for detailed explanations see RKG03).
While the background fields used in that paper are 
derived from the equatorial regions of global dipolar fields, we nevertheless
apply here the corresponding results in the vicinity of the polar cap region.
There, a global poloidal field can no longer
provide a suitable background field structure -- a toroidal field is needed. 
Applicability of the results of RKG03 then requires that the toroidal
background field near the pole and the poloidal one near the equator may have
more or less similar radial profiles. (Note, that the properties of the HDI modes
are only moderately sensitive with respect to the radial profile.) 
This seems to be at 
least possible in the case of convective--dynamo generated fields and even a
reasonable assumption for 
thermoelectrically generated fields with not too low multipolarity. 
For a maximum strength of the toroidal background field in 
the order of about $10^{14}$~G the perturbation rises with a 
characteristic growth time of about $5\cdot10^4$ years.
We assume, that at saturation it may
reach a significant fraction of the background field's strength.
Then it is possible that its surface
strength reaches $5 \cdot 10^{12}$~G and we scaled the perturbation field
(the amplitude of which
is not determined by the linear Eq. (\ref{indeqlin})) simply to this value.

The lifetime of such a perturbation is
bounded from above by its ohmic decay time. Very likely, it is shorter
because the perturbation is subject to the Hall--effect which is
supposed to accelerate the decay in general.
With the conditions at the depth in which the perturbation currents circulate and their scales,
we estimate the ohmic decay time of the field 
presented in Figs.~\ref{eigfunc1} and \ref{eigfunc2} as $\gtrsim 10^7$ yrs.

Both the major radial and tangential scales of the unstable perturbations are 
given approximately by the radial extent of the background field. According to the 
assumptions about the latter they are significant portions of the crust
thickness (say $\lesssim 50$\%) and scale with it. The crust thickness, in turn, is model dependent:
Stiff equations of state (EOS) result in smaller compactnesses and hence larger
star radii and crust thicknesses in comparison with softer ones. For example,
the specific EOSs considered in RKG03, the stiff Pandharipande--Smith
and medium soft Friedman--Pandharipande ones yield a crust thickness of $\sim$ 3.8 km and 700 m, respectively.
Generalising from these two cases
we suggest that the scales of the eigenmodes are not very sensitive
with respect to the star model.
 
Let us stress that the spot--like structure presented in Figs.~\ref{eigfunc1}
and \ref{eigfunc2} satisfies
all conditions to create the vacuum gap and to generate the electron--positron
plasma within it.
In fact, \citet{GM01} and \citet{GM02}
argued that the vacuum gap can only be formed under the so--called
near--threshold conditions, when the surface magnetic field $B_s \ga 0.1 B_q
\sim 4.4 \cdot 10^{12}$~G.
In addition, it requires small radii of curvature $R_c \ll 10^6$~cm.
As can be seen from Fig.~\ref{eigfunc1}, in our case $R_c < 10^5$~cm.
In such a strong and curved surface
magnetic field the magnetic pair creation via curvature and/or ICS (inverse Compton
scattering) photons is very efficient. 

Moreover, the magnetic field lines of the perturbation field converge at
local poles (spot centers)
as can be seen in Fig.~\ref{eigfunc2}.
As already
mentioned at the end of \sectref{obs}, this property of the surface magnetic field
is crucial for the subpulse drift phenomenon.
Of course, as an implicit assumption
one has to keep in mind that just one spot structure plays the role of a
local magnetic pole requiring that the canonical polar cap (i.e., the one formed by 
the dipolar field alone) coincides at least partially with this spot.
%Further on, one must not forget, that only the perturbation field is
%displayed in Figs.~\ref{eigfunc1} and \ref{eigfunc2}. Hence, 
The actual polar cap is
defined by the open field lines of the superimposed global star--centered--dipole
and local perturbation fields, along which an ultra--high
accelerating potential drop can exist, or,
in other words, by those perturbation field lines
which merge with the open dipolar field lines before entering the radio emission
region \citep[see][for details]{GM02a}.

When assuming that close to the surface the global dipolar field is at least
five times smaller than the perturbation field the superposition of both fields
would yield a field very similar to the one shown where only the symmetry
between the regions with positive and negative normal flux is slightly disturbed.  
 
 Note, that the spot--like structures of the surface field must not be confused with the Deshpande--Rankin
 `hot spots' or `sparks' \citep{DR99}. The former can be considered 
 stationary within all the time scales relevant for observations and provide
 just the conditions for the development of sparks.
 By virtue of the $\Evec\times\Bvec$ plasma drift the latter move around the
 local magnetic pole. As the appearance of the pure $\Evec\times\Bvec$ drift
 is confined to homogeneous fields only, additional drift components and/or
 acceleration prevent the charge carriers forming the sparks from moving exactly periodically 
 along closed paths. So it remains to be examined whether
 our field spots are suited to reproduce the periodicities of the subpulse
 phenomenon in detail.

%\enlargethispage{\baselineskip}
\begin{acknowledgements} 
This paper is supported in part by Polish grant 2P03D00819. J.G. acknowledges
the renewal of the Alexander von Humboldt fellowship. U.G. and M.R. are grateful
to the Arbeitsamt Berlin for financial support and to the AIP for hospitality.
\end{acknowledgements} 

\bibliographystyle{aa}

\end{document}